\documentclass[%
a4paper,							
11pt,								
bibliography=totoc,						
abstracton,					
]
{scrartcl}
\usepackage[a4paper,left=3.0cm,right=2.5cm, top=2.5cm, bottom=3.0cm]{geometry}
\usepackage[headsepline=.4pt,footsepline=.4pt,automark,autooneside=false,]{scrlayer-scrpage}
\clearpairofpagestyles
\automark[subsection]{section} 
\automark*[subsubsection]{subsection}
\ihead{\scriptsize\rightmark} 
\usepackage[ngerman, english]{babel}
\usepackage[T1]{fontenc}
\usepackage[utf8]{inputenc}
\usepackage{lmodern} 

\usepackage{calc}
\usepackage{xspace}

\usepackage{amsmath}
\usepackage{amssymb}
\usepackage{amsfonts}
\usepackage{amsthm,thmtools}
\usepackage{mathtools}
\usepackage{bbm}

\theoremstyle{plain}
\newtheorem{theorem}{Theorem}[section]

\theoremstyle{definition}

\theoremstyle{remark}
\newtheorem{remark}[theorem]{Remark}

\usepackage{graphicx,float}
\usepackage[format=plain]{caption}

\usepackage{setspace}

\usepackage{xcolor}

\usepackage{fancyvrb}

\usepackage{paralist}

\usepackage{tabularx}
\usepackage{booktabs}
\usepackage{multirow}
\usepackage{multicol}
\usepackage{pdflscape}
\usepackage{diagbox}

\usepackage[final]{showkeys}

\usepackage[
	backend=bibtex,
	style=authoryear-comp,
	maxbibnames=9,
	maxcitenames=2,
	dashed=false,
	natbib=true,
	sortcites=true,
	block=space
]{biblatex}

\usepackage{orcidlink}
\newcommand{\orcid}[1]{\textsuperscript{\hspace{4pt},\orcidlink{#1}}}
\usepackage{titleref}
\usepackage{hyperref}
\usepackage{cleveref}
\hypersetup{
    colorlinks,
    linkcolor={red},
    citecolor={blue},
    urlcolor={blue}
}

\newenvironment{key}[1][]{ {\noindent \bf Keywords#1: } \rmfamily
}{\hspace*{\fill}  \\}

\newenvironment{code}[1][]{ {\noindent \bf Code availability#1: } \rmfamily
}{\hspace*{\fill}  \\}%

\renewcommand{\P}{\mathbb{P}}
\newcommand{\Q}{\mathbb{Q}}
\newcommand{\E}{\mathbb{E}}
\newcommand{\R}{\mathbb{R}}
\newcommand{\N}{\mathbb{N}}
\DeclareMathOperator*{\argmin}{argmin}

\newcommand{\currency}{NOK}
\newcommand{\conversionRate}{\mathrm{c}}
\newcommand{\FC}{\mathrm{F}}
\newcommand{\cumFC}{\mathrm{CF}}
\newcommand{\HC}{\mathrm{H}}
\newcommand{\cumHC}{\mathrm{CH}}
\newcommand{\numFishes}{\mathrm{n}}
\newcommand{\moratlityRate}{\mathrm{m}}
\newcommand{\biomass}{\mathrm{B}}
\newcommand{\growth}{\mathrm{w}}
\newcommand{\PC}{\mathrm{PC}}

\newcommand{\relImprovement}{\mathrm{RI}}
\newcommand{\fromDate}{01/01/2006\xspace}
\newcommand{\toDate}{01/01/2023\xspace}
\newcommand{\fromDateSub}{01/01/2018\xspace}
\newcommand{\toDateSub}{01/01/2022\xspace}

\newcommand{\CPU}{Intel(R) Core(TM) i9-13900K CPU @ 3.00\,GHz\xspace}
\newcommand{\RAM}{2x32\,GB (Dual Channel) Kingston DIMM DDR5 RAM @ 5600 MHz\xspace}

\newcommand{\OS}{Windows 11 Pro\xspace}

\SaveVerb{python}=(Intel-)Python 3.10=
\newcommand{\python}{\protect\UseVerb{python}\xspace}
\SaveVerb{tensorflow}=Tensorflow 2.8.0=
\newcommand{\tensorflow}{\protect\UseVerb{tensorflow}\xspace}
\SaveVerb{softmax}=softmax=
\newcommand{\softmax}{\protect\UseVerb{softmax}\xspace}
\SaveVerb{relu}=relu=
\newcommand{\relu}{\protect\UseVerb{relu}\xspace}
\SaveVerb{matlab}=Matlab 2023a=
\newcommand{\matlab}{\protect\UseVerb{matlab}\xspace}
\SaveVerb{ga}=ga=

\SaveVerb{fmincon}=fmincon=
\newcommand{\fmincon}{\protect\UseVerb{fmincon}\xspace}
\SaveVerb{fitdist}=fitdist=

\SaveVerb{lsqnonlin}=lsqnonlin=

\newcommand{\matlabGOtoolbox}{(Global) Optimization Toolbox\xspace}

\begin{document}
\onehalfspacing
\pagenumbering{Roman}
\begin{titlepage}

\title{On the Impact of Feeding Cost Risk in Aquaculture Valuation and Decision Making}

\author{Christian Oliver Ewald\thanks{Department of Mathematics and Statistics, Umeå University, Sweden, and Inland University of Applied Sciences, Lillehammer, Norway, e-mail: christian.ewald@umu.se}\orcid{0000-0003-3288-0164} \and Kevin Kamm\thanks{Department of Mathematics and Statistics, Umeå University, Sweden, e-mail: kevin.kamm@umu.se}\orcid{0000-0003-2881-0905}}

\maketitle

\begin{abstract}
\noindent We study the effect of stochastic feeding costs on animal-based commodities with particular focus on aquaculture. More specifically, we use soybean futures to infer on the stochastic behaviour of salmon feed, which we assume to follow a Schwartz-2-factor model. We compare the decision of harvesting salmon using a decision rule assuming either deterministic or stochastic feeding costs, i.e. including feeding cost risk. We identify cases, where accounting for stochastic feeding costs leads to significant improvements as well as cases where deterministic feeding costs are a good enough proxy. Nevertheless, in all of these cases, the newly derived rules show superior performance, while the additional computational costs are negligible. From a methodological point of view, we demonstrate how to use Deep-Neural-Networks to infer on the decision boundary that determines harvesting or continuation, improving on more classical regression-based and curve-fitting methods. To achieve this we use a deep classifier, which not only improves on previous results but also scales well for higher dimensional problems, and in addition mitigates effects due to model uncertainty, which we identify in this article.
\end{abstract}

\medskip
\begin{key}
Deep Learning, Optimal Stopping, Real Options, Commodity Futures, Aquaculture
\end{key}




\medskip
\begin{code}
The Matlab/ Python code and data sets to produce the numerical experiments are available at
\url{https://github.com/kevinkamm/AquacultureStochasticFeeding}.
\end{code}

\thispagestyle{empty}
\end{titlepage}

\newpage
\pagestyle{scrheadings}\ihead{\scriptsize\rightmark}\pagenumbering{arabic}

\section{Introduction}\label{sec:introduction}
In this paper, we extend the results of \citet{Ewald2017} for valuation and optimal decision-making in aquaculture management, taking account of risks attached to a particular input factor: feed. While feed costs have been clearly identified as a potential risk factor for aquaculture production, see for example \citet{Luna2023} and \citet{Misund2022}, feed cost risk has not been taken into account in models for valuation and decision making of aquaculture businesses. This paper is the first to explore this topic. Its relevance is increased due to higher recent volatility in commodity markets. 

 More specifically, we will discuss whether accounting for the possibility of feeding cost risk in valuation and decision-making makes a significant impact, given the current scale of such risk relative to salmon output price risk. This is a very relevant questions for aquaculture businesses and their investors. In doing so, we introduce a methodology for comparing stopping rules from different model frameworks to make a meaningful judgement. First, we demonstrate a comparison of different stopping times, when conditioned on single paths of the underlying stochastic process. Second, we train Deep Classifiers to distinguish between the stopping and continuation region. Both methods will yield very similar results, demonstrating that the neural networks learn the decision rule very well. To this end, we will mainly use the Least-Square-Monte-Carlo regression (LSMC) combined with \citet{Longstaff2001} to perform the backward induction in the optimal stopping problem for harvesting the salmons. We also implemented a Deep Learning approach like in \citet{Becker2021} and observed that it performs well for this purpose.
 
The focus of this article is to understand the impact of feeding costs risk on the decision-making in harvesting salmon. According to \citet[pp.~28\,ff.]{Misund2022} a typical blend of salmon feed consists of soybean meal (30--44 percent), soybean oil, rapeseed oil, wheat, corn, fish meal, and fish oil. The soy-based ingredients make up the majority of the blend and for this reason, soybean futures markets provide a good proxy for future price development of salmon feed and its risk-characteristics, as well as providing hedging opportunities for aquaculture businesses.

We will model both commodities, salmon and soy, using two decoupled Schwartz-2-factor models (cf. \citet{Schwartz1997}). We will investigate the impact of parameters on the harvesting decision, as well as problems concerning the estimation of the model parameters. More specifically, we will compare two different techniques to calibrate the models: The first uses a Kalman filter as originally suggested by \citet{Schwartz1997}, it is being facilitated by a large number of authors. The second one uses a nested minimization based on \citet{Cortazar2003}, and is facilitated by a somehow smaller number of authors. Both approaches are conceptually very different. We find that both methods capture the unobservable state variables well, but lead to very different estimates for 
 the model parameters. Both of these different parameter sets however result in a relatively good fit to the futures market data. For pricing futures this discrepancy may therefore be less of an issue, but, as we demonstrate, it has a significant impact on the real option valuation and the optimal harvesting rule discussed in this article. Such model uncertainty seems to be neglected in the literature and requires more attention in future investigations of real options.

To assume that the two commodities salmon and soy move independently from each other may seem crude.
In fact from the modeling perspective, this assumption could be easily dropped, and the optimal stopping problem could still be solved in the way we propose here. However, for the parameter estimation, it is a bigger issue to include inter-commodity correlations reliably. We studied the market data of both commodities and found different results for their correlation. For data further in the past, we estimated a small (negligible) correlation, while more recently a more significant correlation can be found. However, this might be because of a common factor, most likely inflation, and will not be further pursued here. In this paper, we will focus on stochastic feeding costs in different scenarios, as described further below.


For an overview of the historical development for salmon commodity pricing and valuation models, we refer the reader to \citet{Ewald2017}. Additionally, 
for a comprehensive treatment of all the economic factors of fish farming we refer the reader to \citet{Misund2022} and \citet{Luna2023}. In this article, we focus on (stochastic) feeding costs.

\citet{Shepherd2017} discuss salmon feed in detail and consider supply-chain issues, which are implicitly included in this paper, since in the event of supply shortage, we would expect the future prices to rise accordingly. Considering supply-chains, we could also look at storage models such as in \citet{Osmundsen2021}. \citet{Gomes2023} investigate salmon feed intake and overfeeding, which could be linked to a storage model.  

Nevertheless, our article is the first article that looks at the impact of feed cost risk on the decision-making of aquaculture businesses. There is nothing alike in the literature. 

\medskip
The remainder of this paper is structured as follows:
In \Cref{sec:framework}, we introduce the underlying commodity price model, followed in \Cref{sec:salmonFarmParameters} by a description of the salmon farm features considered in this paper. Afterwards, in \Cref{sec:stopping}, we formulate the optimal decision making problem using both deterministic and stochastic feeding costs.
Numerical results will be discussed in \Cref{sec:numerics}. We first describe the market data in \Cref{sec:marketData}, followed by two calibration algorithms in \Cref{sec:calibration}. Following this, we explain our methodology for comparing stochastic and deterministic feeding rules in \Cref{sec:stoppingBoundaries}. For this we involve a deep neural network. Our main results are presented and discussed in \Cref{sec:results} while the main conclusions of this paper are summarized in \Cref{sec:conclusion}.

\section{Mathematical Model and Framework}\label{sec:framework}
Henceforth, let $\left(\Omega,\mathcal{F},\Q\right)$ be a probability space and $\Q$ be a risk-neutral measure. Moreover, let $T>0$ be a finite time-horizon and $r>0$ a fixed deterministic interest rate.

We will use a multi-commodity framework consisting of two independent Schwartz-2-factor models directly under $\Q$, i.e.
\begin{align*}
    dS_t^i &= \left(r-\delta_t^i\right) S_t^i dt + \sigma^i_1 S_t^i dW_t^{i,1}, \quad S_0^i=s_0^i \in \R_{\geq 0,}\\
    d\delta_t^i &= \left(\kappa^i\left(\alpha^i -\delta^i_t\right) -\lambda^i\right) dt + \sigma_2^i dW_t^{i,2},  \quad \delta_0^i=d_0^i \in \R\\
    d\langle W^{i,1},W^{i,2}\rangle_t&= \rho^i dt,\qquad d\langle W^{i,1},W^{j,2}\rangle_t=0,\quad i\neq j,
\end{align*}
where $W_t\coloneqq \left(W_t^{i,j}\right)_{i,j=1,2}$ are Brownian motions generating the $\sigma$-algebra $\mathcal{F}_t$ augmented by $\Q$ nullsets satisfying the usual conditions.

The dynamics $dS_t^i$ describe the i-th commodity's spot price with convenience yield described by $d\delta_t^i$. The parameter $\sigma_1^i>0$ is the spot volatility, $\kappa^i>0$
the mean reversion speed of the convenience yield, $\alpha^i\in \R$ the long-term mean, $\lambda^i\geq 0$ a risk-premium and $\sigma_2^i>0$ the volatility of the convenience yield.

It is straightforward to implement the optimal stopping problem in \Cref{sec:stopping} so as to include inter-correlations for the Brownian motions or factors for coupling of the SDEs. However, this is not the focus of this paper and a study of such effects is deferred to future research.\footnote{The coupling of the SDEs will pose some difficulties in the parameter calibration using standard tools like Kalman-filtering, since closed form solutions for future prices will not be available anymore. Inter-correlations might also be difficult to estimate from the market data and may vary over time or exhibit stochastic behaviour.}

Henceforth, $S^1_t$ will take the role of the salmon spot price and $S_t^2$ will take the role of the soybean spot price.
Since we assume independence between $S^1_t$ and $S^2_t$, standard tools for calibration of both commodities can be applied individually and is referred to \Cref{sec:calibration}.
Additionally, since we assume that the commodities are uncoupled, we can use the results in \citet{Schwartz1997} to obtain an explicit formula of the future/ forward\footnote{We assume that our interest rate is deterministic and constant. Therefore, forward and future prices are the same, cf. \citet[p.~456 Proposition 29.6]{Bjork2009}.} prices in this model. We have for given spot price $S\in \R$, convenience yield $\delta\in \R$ and time-to-maturity $\Delta T\geq 0$
\begin{align*}
    F(S,\delta,\Delta T)&=
    S \exp\left( 
        -\delta \frac{1-e^{-\kappa \Delta T}}{\kappa} + A(\Delta T)
    \right)\\
    A(\Delta T)&=
    \left(
        r - \hat{\alpha} + \frac{1}{2}\frac{\sigma_2^2}{\kappa^2}-\frac{\sigma_1 \sigma_2}{\kappa}
    \right)\Delta T +
    \frac{1}{4} \sigma_2^2 \frac{1-e^{-2\kappa \Delta T}}{\kappa^3}+
    \left(
        \hat{\alpha} \kappa + \sigma_1 \sigma_2 \rho - \frac{\sigma_2^2}{\kappa}
    \right)\frac{1-e^{-\kappa \Delta T}}{\kappa^2},
\end{align*}
where $\hat{\alpha}\coloneqq \alpha - \frac{\lambda}{\kappa}$ , omitting superscripts $i$ referring to the $i$-th commodity for ease of notation, thus superscripts have to be understood as exponents.
\subsection{Salmon Farm Parameters}\label{sec:salmonFarmParameters}
In this subsection, we briefly describe the features of a typical salmon farm, referring the reader to \citet{Misund2022} for a detailed discussion on general economic issues concerning salmon farms.

We consider a single farm over one harvesting cycle, hence financial values correspond to the value of a lease, as discussed in \citet{Ewald2017}. This means that a fish farmer leases the farm, buys $\numFishes_0\in \N$ smolt, young salmon, and feeds them till they are ready for harvest, and at time of harvest sells the fish for the market price and returns the farm. The growth of the fishes over time depends on various factors like water temperature, amount and quality of feed, health, etc. We simplify this by considering a deterministic function over time, as so called \emph{Bertalanffy’s growth function}, which is given by 
\begin{align*}
    \growth(t)&\coloneqq \growth_\infty \left(a-b\, e^{-ct}\right)^3,
\end{align*}
for parameters $a,b,c$ given in \Cref{tab:fishPoolParameters}. This function measures the growth in kg per fish. We assume a constant mortality rate $\moratlityRate>0$ and the total number of fishes $\numFishes(t)$ at time $t$ follows the deterministic dynamics 
\begin{align*}
    d \numFishes(t) &= -\moratlityRate\, \numFishes(t) dt, \quad \numFishes(0)=\numFishes_0.
\end{align*}
Now, we are able to measure the total biomass (in kg) of the fish farm over time by setting
\begin{align*}
    \biomass(t) = \numFishes(t) \growth(t).
\end{align*}

We will consider only two varying factors of production costs: harvesting costs and feeding costs, all other costs including labor, medical treatments, capital costs, etc., will be treated as constants for simplicity. Harvesting costs are given by $\HC_0$ per kg of fish, the total harvesting costs of the fish farm at time $t>0$ are therefore
\begin{align*}
    \cumHC(t)=\HC_0\, \biomass(t).    
\end{align*}

Feeding costs will be the main focus of this paper. We use a conversion rate of how much kg of feed will convert to kg of fish 
\begin{align*}
    \conversionRate \coloneqq 
    1.1\frac{\text{\,kg feed}}{\text{\,kg fish}},
\end{align*}
a reasonable value. Let $F(0)=F_0$ be the initial feeding cost for one fish per year. We will infer the feeding costs from the relative changes of soybean prices $S_t^2$, i.e.
\begin{align*}
    \tilde{S}_t^2 \coloneqq \frac{S_t^2}{S_0^2},
\end{align*}
by using
\begin{align*}
    \FC^\text{stoch}_t = \FC_0 \tilde{S}_t^2, \quad \FC^\text{determ}_t = \FC_0 \E\left[\tilde{S}_t^2\right],
\end{align*}
and define the discounted cumulative total feeding costs as\footnote{The time-derivative of the growth is denoted by $\partial_t \growth$ and given by $\left(\partial_t \growth\right)(t)= 3\growth_\infty \left(a-b\, e^{-ct}\right)^2\, b\,c\, e^{-ct}$. The term $\left(\partial_t \growth\right)(s)$ captures the weight increase of one fish in kg, $\left(\partial_t \growth\right)(s)\, \conversionRate$ the amount of feed in kg that is required for this growth, $\numFishes(s) \left(\partial_t \growth\right)(s)\, \conversionRate$ the total feed and finally $\FC_s \, \numFishes(s) \left(\partial_t \growth\right)(s)\, \conversionRate$ the total cost.} 
\begin{align*}
    \cumFC(t) = \int_0^t{ e^{-rt} \left(\FC_s \, \numFishes(s) \left(\partial_t \growth\right)(s)\, \conversionRate\right)  ds}
\end{align*}
for general feeding costs $\FC_s$, where once more an additional upper index referring to stochastic vs. deterministic is omitted for ease of notation.

The parameters in \Cref{tab:fishPoolParameters} are mostly taken from \citet{Ewald2017} and references therein. The initial feeding and harvesting costs are estimated from \citet[p.~25 Figure 9]{Misund2022}. We will use the values in \Cref{tab:fishPoolParameters} for the remainder of this paper.
\begin{table}[]
    \centering
    \caption{Fish farm parameters.}
    \begin{tabular}{l*{3}{c}}
        Name                        & Symbol            & Value                                 & Unit\\
        \toprule
        Bertalanffy growth factor   & $a$               & 1.113                                 &  -     \\    
        Bertalanffy growth factor   & $b$               & 1.097                                 &  -     \\ 
        Bertalanffy growth factor   & $c$               & 1.43                                  &  -     \\ 
        Asymptotic weight           & $\growth_\infty$  & 6                                     & kg    \\ 
        Mortality rate              & $\moratlityRate$  & 10                                    & \%    \\
        Conversion rate             & $\conversionRate$ & 1.1                                   & kg feed/ kg fish\\
        Number of recruits          & $\numFishes_0$    & 10000                                 & fish \\
        Time horizon                & $T$               & 3                                     & years\\
        Harvesting possibilities    & $N$               & 72                                    & -\\
        Salmon spot price           & $\hat{S}_0^1$     & see \Cref{tab:paramSalmon}            & \currency/ kg\\
        Production costs            & $\PC$             & $0.5\, \hat{S}_0^1$                   & \currency/ kg\\
        Harvesting costs            & $\HC_0$           & $0.1\, \PC$                           & \currency / kg\\
        Feeding costs               & $\FC_0$           & $0.25\, \PC$                          & \currency / kg year\\
        Salmon initial value        & $S_0^1$           & $\hat{S}_0^1 -\PC + \HC_0 + \FC_0$    & \currency/ kg\\
        Soy initial value           & $S_0^2$           & 1                                     & - \\
    \end{tabular}
    \label{tab:fishPoolParameters}
\end{table}

\subsection{Optimal Stopping Problem}\label{sec:stopping}
Following \citet[pp.~8\,ff.]{Ewald2017}, the objective of a fish farmer is to find the optimal harvesting time for salmon cultivated on the farm, where optimal has to be understood as maximizing the expected value under a risk-neutral measure. The latter corresponds to maximizing the financial value of the farm, when appropriately taking account of relevant risk premia, see \citet{Ewald2017}. We consider the current value of the fish at their current weight $S_t^1 \biomass(t)$ minus the harvesting costs at $\cumHC(t)$ and the cumulative feeding costs up to this point in time $\cumFC(t)$. Let $X_t^\text{stoch}\coloneqq \left(S_t^1,\delta_t^1,S_t^2,\delta_t^2\right)$ and
$X_t^\text{determ}\coloneqq \left(S_t^1,\delta_t^1\right)$. The optimal stopping problem for stochastic and deterministic feeding costs becomes respectively
\begin{align}
    W_0^\text{stoch}(x)&\coloneqq
    \sup_{\tau^\text{stoch}}
        \E^\Q\left[\left.
            \exp\left(- r \tau\right)
            \left(
                S_\tau^1 \biomass(\tau) - \cumHC(\tau)
            \right)
             - \cumFC^\text{stoch}(\tau)
             \right|
             X_0^\text{stoch}=x
        \right],\\
    W_0^\text{determ}(x)&\coloneqq
    \sup_{\tau^\text{determ}}
        \E^\Q\left[\left.
            \exp\left(- r \tau\right)
            \left(
                S_\tau^1 \biomass(\tau) - \cumHC(\tau)
            \right)
             - \cumFC^\text{determ}(\tau)
             \right|
             X_0^\text{determ}=x
        \right].
        \label{eq:optStopping}
\end{align}
In this paper, we compare the stopping rules obtained from $W_0^\text{stoch}(x)$ and $W_0^\text{determ}(x)$ by evaluating
\begin{align}
 V_0^z(x)\coloneqq \E^\Q\left[\left.
    \exp\left(- r \tau^z\right)
    \left(
        S_{\tau^z}^1 \biomass(\tau^z) - \cumHC(\tau^z)
    \right)
     - \cumFC^\text{stoch}(\tau^z)
     \right|
     X_0^\text{stoch}=x
     \label{eq:objective}
\right]   
\end{align}
for $z \in \{\text{stoch},\,\text{determ}\}$ in an appropriate way as described in \Cref{sec:stoppingBoundaries}, and investigate the question whether $V_0^\text{determ}(x)$ is a good enough approximation of $W_0^\text{stoch}(x)$
or alternatively in which cases it is beneficial to consider the slightly more complicated stopping rule $\tau^\text{stoch}$.

We will solve both of the optimal stopping problems numerically by using least-square Monte Carlo (LSMC), i.e. a backward induction with the \citet{Longstaff2001} algorithm and refer the reader to \citet[pp.~9\,ff.]{Ewald2017} for the algorithmic details. We use polynomials up to power two for the regression of the conditional expectations in our implementation.

\begin{remark}\label{rem:deepOptimalStopping}
It is straightforward to use the Deep Optimal Stopping Network by \citet{Becker2021} and we provide the code for using this as well. 
This allows in principle for high-dimensional optimal stopping problems, which could become relevant if one considers further stochastic factors. For example, using soybean futures as a surrogate for the feeding costs is a more or less crude approximation of reality, the feed consists of several different ingredients, also including maize and fish meal (cf. \citet[p.~29 Figure 11]{Misund2022}), all could be modeled stochastically as well. Introducing stochastic mortality and stochastic growth of the fish would add to the further dimensions, making a least-square Monte-Carlo regression less and less appealing, as the computational costs increase, while the deep neural networks copes well.    
\end{remark}

\section{Numerical Experiments}\label{sec:numerics}
In this section, we will carry out some numerical experiments. In \Cref{sec:marketData}, we will briefly discuss the historical data used for the calibration process as described in \Cref{sec:calibration}. After the calibration, we will compare the different feeding cost scenarios. We introduce two methods for this comparison in \Cref{sec:stoppingBoundaries}. The first method compares the stopping rules obtained from LSMC pathwise and the second method relies on classifying the stopping region with a Deep-Neural-Network (DNN), providing an excellent alternative to the more classical curve fitting process to identify the optimal exercise boundary, as employed in \citet{Ewald2017} for example. Lastly, in \Cref{sec:results}, we present our results and draw our main conclusion: In fact, it is beneficial to consider stochastic feeding costs, at least under certain scenarios which we identify.

For the calibration we used \matlab with the \matlabGOtoolbox
and for the DNN \python with \tensorflow
running on \OS, on a machine with the following specifications: processor
\CPU and \RAM. A GPU did not improve the performance in this case.

For all of our experiments we will fix the interest rate to $r=0.0303$.
\subsection{Market Data}\label{sec:marketData}
\begin{figure}
    \centering
    \includegraphics[width=.49\columnwidth]{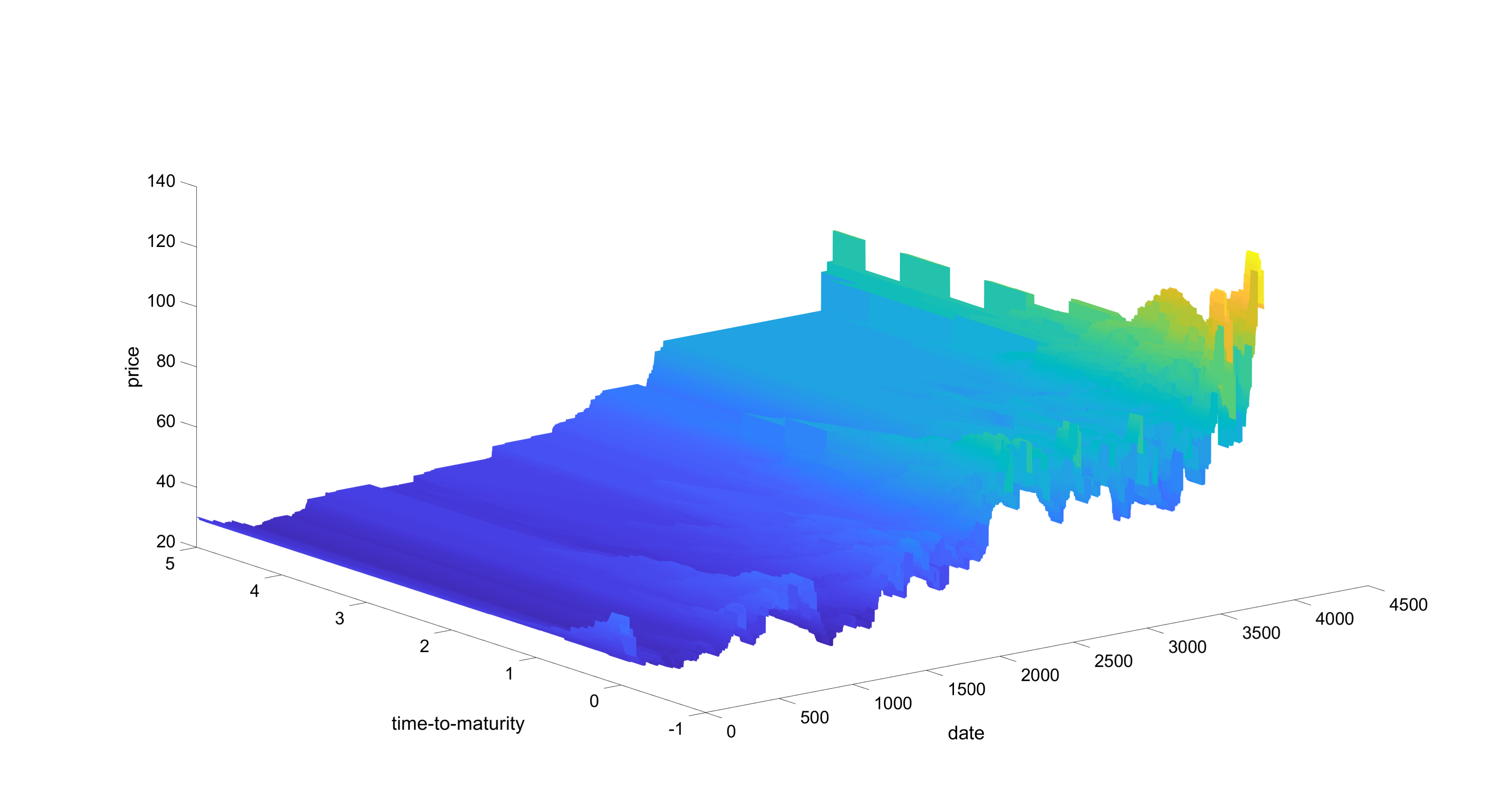}
    \includegraphics[width=.49\columnwidth]{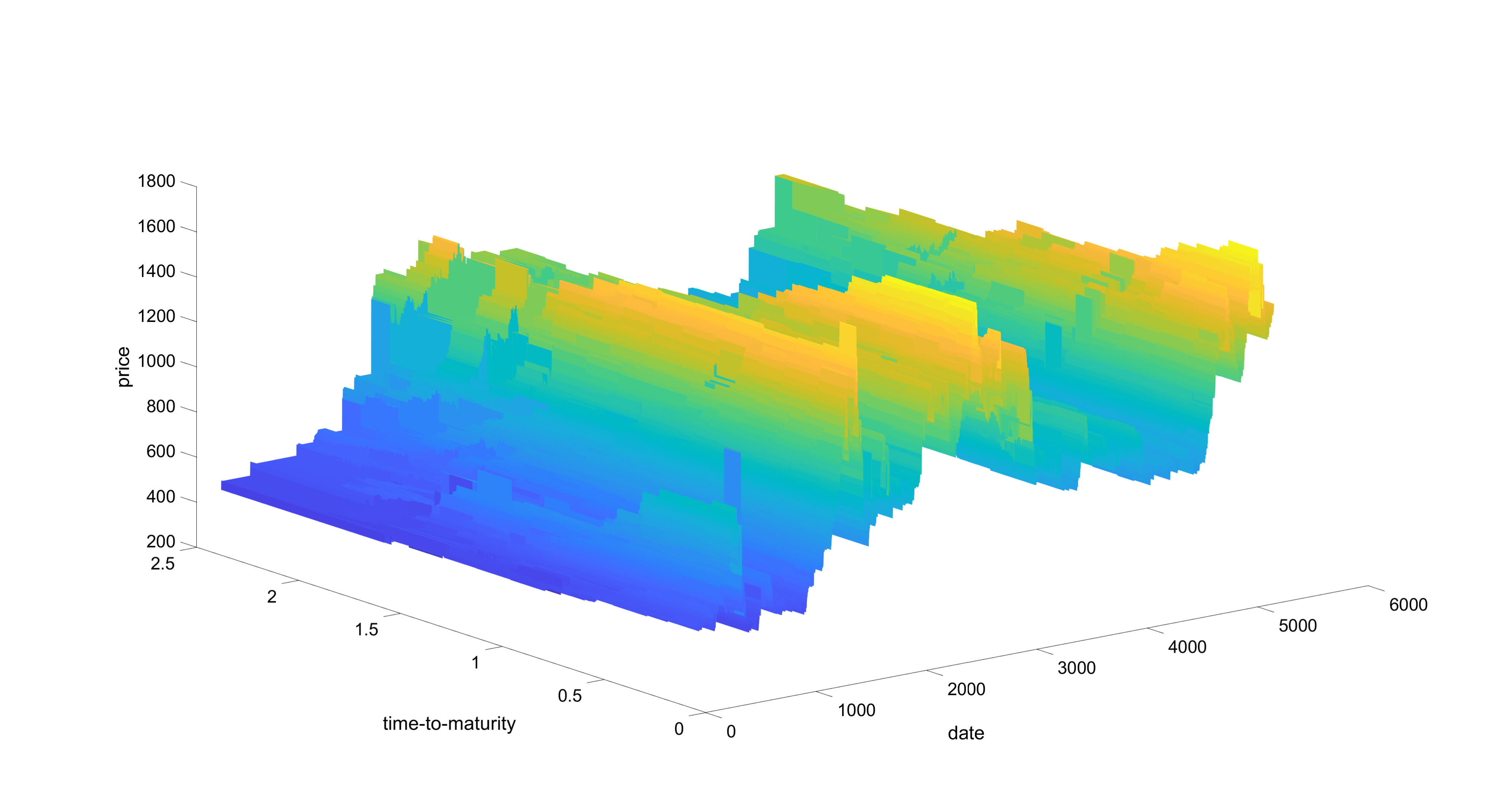}
    \caption{Interpolated (nearest neighbour) market future surfaces for salmon (left) and soy (right) from \fromDate till \toDate. On the $x$-axis are dates identified by integers, on the $y$-axis the time-to-maturity and on the $z$-axis the corresponding price. Deep blue associates a low and yellow a high price.}
    \label{fig:marketSurface}
\end{figure}

We use historical future data from \fromDate till \toDate for both salmon\footnote{The historical salmon future prices are publicly available on\\ \url{https://fishpool.eu/forward-price-history/} (last accessed 04/07/2023 12:45 CEST).} and soybeans. This is illustrated in \Cref{fig:marketSurface}.
For the plots we used a nearest neighbour interpolation for fixed date in the direction of the time-to-maturity.
A dark blue color reflects low prices, while a bright yellow color reflects high prices relative to the individual commodity.
On the left-hand side we can see the evolution of the salmon future prices over time for different time-to-maturities and on the right-hand side for the soybean futures. Notably, the volatility of these prices seem to increase significantly at the end during the Covid-pandemic. For salmon the prices seem to increase steadily over time with more rapid increase recently. If we estimate the spot prices from the smallest time-to-maturity in these pictures, we can also observe that the majority of the behaviour of the future prices is determined by the spots, since they appear almost constant or with non-rapid fluctuations on the time-to-maturity axis for a fixed date.

\subsection{Calibration to Market Data}\label{sec:calibration}
At the start of this subsection, we would like to remind the reader about two well-known techniques for the calibration of multi-factor commodity pricing models with hidden stare variables
to historical futures price data. Since our commodity models are uncoupled we will omit once again the dependence on the individual commodities throughout this entire section.
The first was introduced by \citet{Cortazar2003} as a nested least square regression the sefcond one
is using a Kalman filter (cf. \citet{Schwartz1997}).

The goal is to find both, the model parameters
\begin{align*}
    \Pi \coloneqq  [\sigma_1,\ \sigma_2,\ \kappa,\ \alpha,\ \lambda,\ \rho],
\end{align*}
and the unobservable spots and convenience yields for each date $S\geq 0$, $\delta \in \R$.

For the calibration procedure we will restrict ourselves to a subset of the main dataset. In fact we only consider the slice from \fromDateSub to
\toDateSub, taking only the most recent price developments into account making a regime-switching model unnecessary.

We will denote in this subsection the number of days, where market future prices are available by $K\in\N$ and for each date the number of available maturities by $P_i\in\N$.
The market future prices at date $i=1,\dots,K$ with time-to-maturity $\Delta T_j$, $j=1,\dots,P_i$ will be denoted by $F^\text{Market}_{ij}$ and the model future prices depending on the parameters $\Pi$ by $F^\Pi$.
\paragraph*{\citet{Cortazar2003}.}
We would like to minimize the following objective
\begin{align*}
    \min_{\Pi} \sum_{i=1}^{K}\sum_{j=1}^{P_i} \left( \log F^\Pi\left(S_{t_i},\delta_{t_i},T_j - t_i\right) - \log F^\text{Market}_{ij}\right)^2
\end{align*}
subject to
\begin{align*}
    (S_{t_i},\delta_{t_i}) \in \argmin_{S,\delta}  \sum_{j=1}^{P_i}  \left( \log F^\Pi\left(S,\delta,T_j - t_i\right) - \log F^\text{Market}_{ij}\right)^2 \text{ for all } i=1,\dots,N.
\end{align*}
We need the constraint since the spot price and convenience yield are not observable.
The optimization for the state variables $S_{t_i},\delta_{t_i}$ can be solved efficiently by a linear least-square regression and for the objective we use \matlab's \fmincon with the interior-point method. This is a simplified version of the algorithm presented in \citet{Cortazar2003} and is only used in order to stress that the model future prices are mostly determined by the unobservable state variables and can have an equally good fit to the market future prices for very different parameters $\Pi$.

\paragraph*{Kalman filtering.}
Following \citet[pp.~10\,ff.]{Schwartz1997}, the idea is to first discretize the log-spot price $s_t\coloneqq \log(S_t)$ and convenience yield under the historical measure $\P$, e.g. by using an Euler-Maruyama scheme. They are both normally distributed. Thus, for a homogeneous time grid $(t_i)_{i=1,\dots,K}$ with step-size $\Delta t$, we have
\begin{align*} 
    s_{t_{i+1}}&=s_{t_{i}} + (\mu-\delta_{t_{i}} -\frac{1}{2}\sigma_1^2) \Delta t + \sigma_1 \Delta W_{t_i}^1,\\
    \delta_{t_{i+1}} &= \delta_{t_{i}} + \left(\kappa \left(\alpha - \delta_{t_{i}} \right) -\lambda \right) \Delta t + \sigma_2 \Delta W_{t_i}^2.
\end{align*}
Letting $X_{t_i}\coloneqq [s_{t_i},\delta_{t_i}]^\top$ we have the \emph{transition equation}
\begin{gather*}
    X_{t_{i+1}} = c + T \cdot X_{t_{i}} + \eta_i,\\
    c\coloneqq\left[\begin{array}{c}
         r - \frac{1}{2}\sigma_1^2  \\
         \kappa \alpha 
    \end{array}\right] \Delta t,\quad 
    T\coloneqq\left[\begin{array}{cc}
       1  & -\Delta t \\
       0  & 1-\kappa \Delta t
    \end{array}\right],\quad 
    E\coloneqq \left[\begin{array}{cc}
       \sigma_1^2  & \rho \sigma_1 \sigma_2 \\
       \rho \sigma_1 \sigma_2  & \sigma_2^2
    \end{array}\right] \Delta t
\end{gather*}
in a so-called state-space formulation. Here $\eta_i\sim\mathcal{N}(0,E)$ is serially uncorrelated. Since the spot prices are non-observable, we need to \emph{measure} them. For this we use the log-future prices
with different time-to-maturities $\Delta T_j$ to derive the \emph{measurement equation}
\begin{gather*}
    Y_{t_{i}} = \left[d(\Delta T_j)\right]_{j=1,\dots,P} + \left[Z(\Delta T_j)\right]_{j=1,\dots,P} \cdot X_{t_{i}} + \epsilon_i,\quad j=1,\dots,P,\ P\in\N,\\
    d(\Delta T)\coloneqq A(\Delta T),\quad 
    Z(\Delta T)\coloneqq\left[\begin{array}{cc}
       1  & \frac{-(1-e^{-\kappa \Delta t}}{\kappa}) 
    \end{array}\right],\quad 
    D\coloneqq \mathrm{diag}\left(d_1^2,\dots,d_P^2 \right),
\end{gather*}
where $\epsilon_i\sim\mathcal{N}(0,D)$ is serially uncorrelated and independent of $\eta$. Notice, that $\epsilon$ has to be added as measurement noise to make the model compatible for Kalman-filtering. The choice of the covariance-matrix has a significant impact on the estimation. We chose in our implementation a diagonal covariance matrix $D$. This choice seems to be the standard for commodity models and has already been used by \citet{Schwartz1997}.

Using different panels, i.e., a collection of different time-to-maturities $\Delta T_j$, can yield significantly different parameters. The panels are assumed to be fixed throughout the entire filtering, which is a disadvantage compared to \citet{Cortazar2003}, since it can use the entire data set without any interpolation or averaging.
In order to compare the Kalman filter to the method by \citet{Cortazar2003}, we use a daily time step for the Kalman filter as well.

The parameters can now be estimated by maximizing the likelihood of $Y_{t_i}$ equalling the market log-future prices and meanwhile estimating the unobservable spot prices and convenience yields by the Kalman-filter. For more details we refer to the code and the monographs \citet[pp.~100\,ff. Chapter 3]{Harvey1989} and \citet[pp.~635\,ff. Chapter 13.3]{Bishop2006}. We use \matlab's \fmincon with the interior-point method to minimize the negative log-likelihood.

\paragraph*{Model uncertainty.}
In \Cref{fig:kalmanVsCortazar}, we compare some calibration results using the soybean data from \fromDateSub till \toDateSub using both a Kalman filter and \citet{Cortazar2003}.
We found the following parameters 
\begin{align*}
    \Pi^\text{Kalman}&= [0.2044,\ 0.1275,\ 0.1541,\ 0.3355,\ 0.0101,\ 0.9065],\\
    \Pi^\text{CS}&= [2.1975,\    0.4594,\    0.4381,\    2.2825,\    1.3120,\     0.4893].
\end{align*}
The red color will show all the results using $\Pi^\text{Kalman}$ and the magenta color for $\Pi^\text{CS}$.
In the upper third of the figure we compare the model future prices to the market future prices for different time-to-maturity on the 16/12/2022, which was randomly chosen. We can see that both models seem to regress the market prices (blue crosses) appropriately and they are very close to each other. In the middle part of the figure we compare the inferred spot prices $S_0$ for each date from \fromDateSub till \toDateSub. We can also see that both calibration methods seem to find very similar spot prices. A little bit more variation can be seen in the inferred convenience yields in the lower part of the figure, but still very similar. We would like to stress at this point, that the model seems to produce very similar results with extremely different parameters. If one only needs to find the unobservable spot prices or fit the model to futures prices, then both methods perform equally well, but if one needs the underlying model for more complex decision making, this raises a major concern, as in our case. In fact, for the real option discussed in this paper, the model parameters will have a significant impact on the harvesting decision. We will address this problem further in future research.
\begin{figure}
    \centering
    \includegraphics[width=\columnwidth]{Figures/KalmanVsCortazar.pdf}
    \caption{Close spot and future prices for different model parameters.}
    \label{fig:kalmanVsCortazar}
\end{figure}

\begin{figure}
    \centering
    \includegraphics[width=.49\columnwidth]{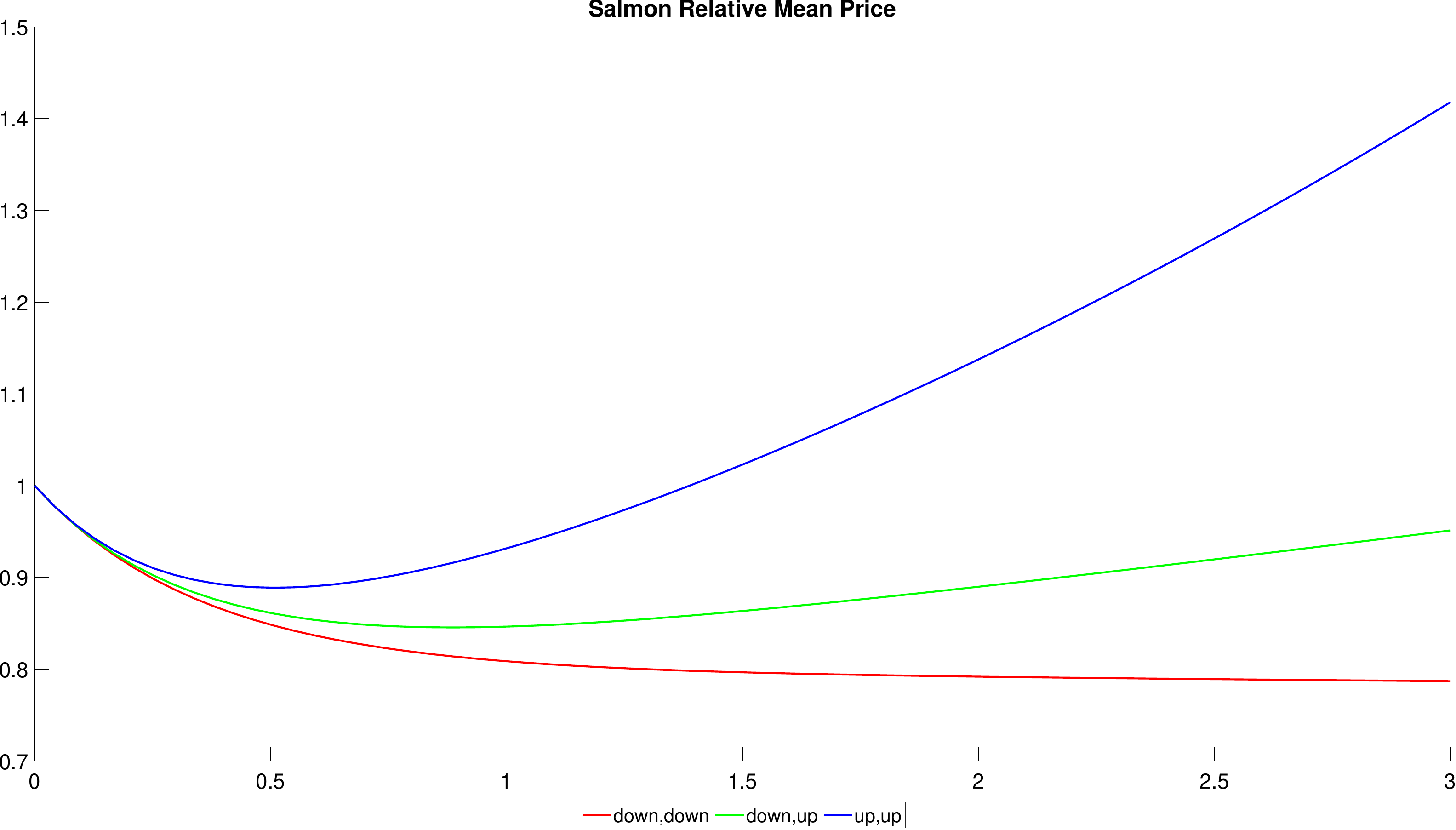}
    \includegraphics[width=.49\columnwidth]{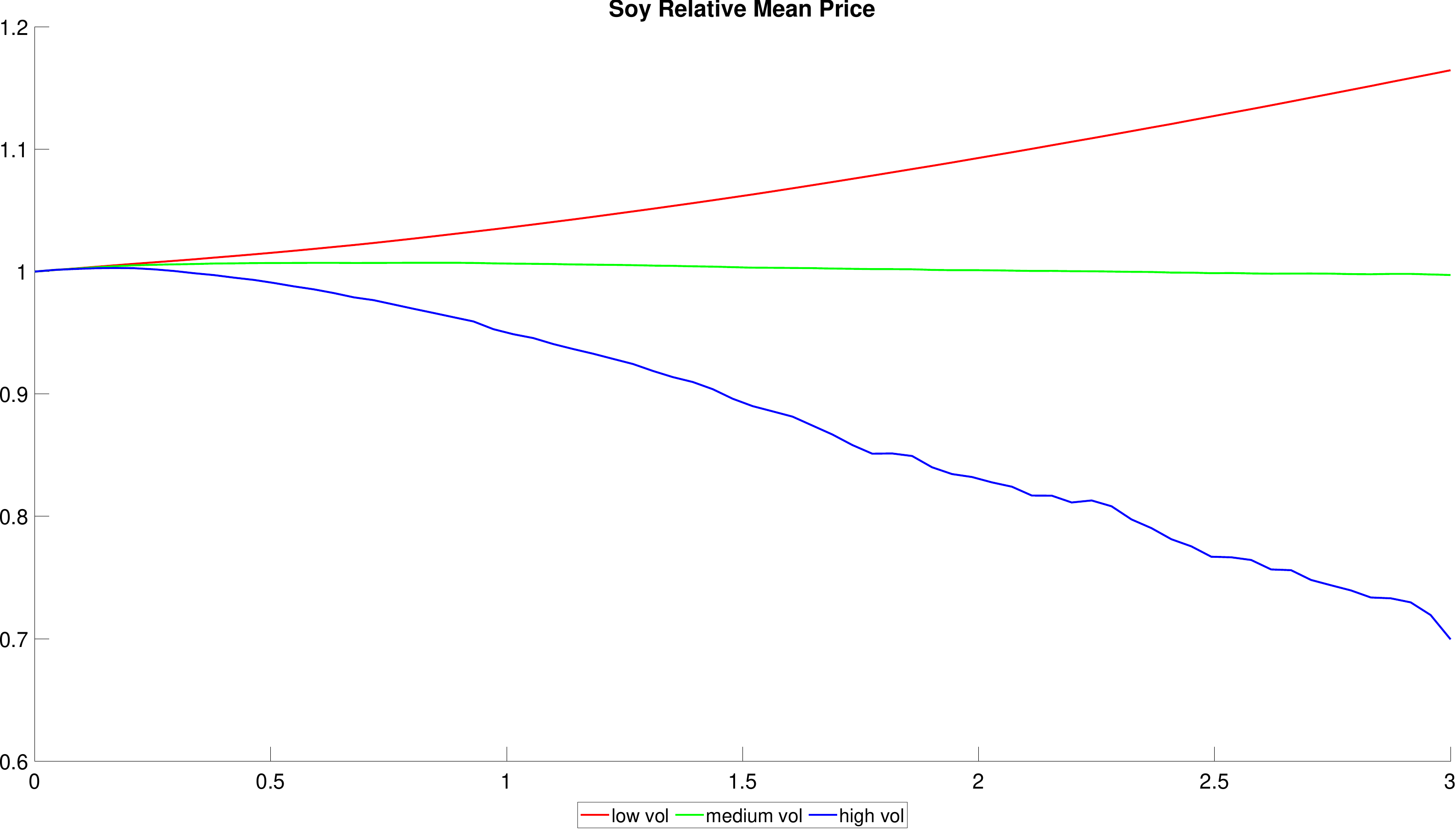}
    \caption{Mean relative changes of salmon (left) and soy (right) prices for the different scenarios in \Cref{tab:paramSalmon}--\ref{tab:paramSoy}.}
    \label{fig:meanSalmonSoy}
\end{figure}

We believe that the Kalman filter underestimates the real volatility, while the method by \citet{Cortazar2003} overestimates it.
Since there seems to be a lot of uncertainty in finding appropriate parameters for our commodity models, we will use the artificial parameters shown in \Cref{tab:paramSalmon} and \Cref{tab:paramSoy} for our further investigation. These are based on the calibration results, and chosen such that we can discuss the question of whether stochastic feeding costs should be considered in aquaculture investment or whether the effect is negligible. The mean relative price changes of the commodity models using the different parameters are illustrated in \Cref{fig:meanSalmonSoy}. The left subfigure shows the salmon prices. The red line corresponds to the scenario \emph{down, down}, the green line to \emph{down, up}, and the blue line to \emph{up, up}. Similarly for the soy spot prices in the right subfigure. The red line corresponds to the scenario \emph{low volatility}, the green line corresponds to \emph{medium volatility} and the blue line corresponds to \emph{high volatility}.

We did a lot of tests and found that the most sensitive parameter for estimating the benefit of using stochastic feeding costs is $\sigma_1^2$ for a given salmon scenario. Moreover, the closer the feeding costs are to the salmon revenue, the more the feeding costs matter, which explains our choices for selecting the salmon scenarios such that we have a mean decrease, mean stable and mean increase scenario in prices. This could be achieved by only adjusting the risk-premium parameter $\lambda^1$, since it changes the drift of the convenience yield and thus the drift of the spot price.

\begin{table}[]
    \centering
    \caption{Parameters for different scenarios of salmon models.}
    \begin{tabular}{l*{8}{c}}
       Scenario          & $\sigma_1^1$ & $\sigma_2^1 $ & $\kappa^1$ & $\alpha^1$ & $\lambda^1$ & $\rho^1$  & $\delta^1_0$ & $\hat{S}_0^1$\\
       \toprule
       down, down        & 0.23         & 0.75          & 2.6        & 0.02       & 0.01        & 0.9       & 0.57         & 95\\
       down, up          & 0.23         & 0.75          & 2.6        & 0.02       & 0.2         & 0.9       & 0.57         & 95\\
       up, up            & 0.23         & 0.75          & 2.6        & 0.02       & 0.6         & 0.9       & 0.57         & 95\\
    \end{tabular}
    \label{tab:paramSalmon}
\end{table}

\begin{table}[]
    \centering
    \caption{Parameters for different scenarios of soybean models.}
    \begin{tabular}{l*{8}{c}}
       Scenario          & $\sigma_1^2$ & $\sigma_2^2 $ & $\kappa^2$ & $\alpha^2$ & $\lambda^2$ & $\rho^2$  & $\delta^2_0$ & $\hat{S}_0^2$\\
       \toprule
       low volatility    & 0.5          & 0.4           & 1.2        & 0.06       & 0.14        & 0.44      & 0            & 1500\\
       medium volatility & 1            & 0.4           & 1.2        & 0.06       & 0.14        & 0.44      & 0            & 1500\\
       high volatility   & 2            & 0.4           & 1.2        & 0.06       & 0.14        & 0.44      & 0            & 1500\\
    \end{tabular}
    \label{tab:paramSoy}
\end{table}

We have made the code publicly available and encourage the reader to try other parameter configurations.

\subsection{Comparison of Exercise Decisions}\label{sec:stoppingBoundaries}
In order to compare the exercise decisions in the different feeding cost scenarios, we will employ two different methods.
The first one, will be a pathwise comparison of the the stopping times and the second one will use a Deep-Neural-Network in order to learn the decision boundary.
\paragraph*{Pathwise Comparison of Stopping Times.}
Let $\left(\Omega,\mathcal{F},\mathcal{F}_t,\P\right)$ be a filtered probability space satisfying the usual conditions, where
$\mathcal{F}_t\coloneqq \sigma\left(X_t\right)$ is generated by a stochastic process, and
$\mathcal{F}^i_t \subseteq \mathcal{F}_t$, $i=1,2$, be two sub-$\sigma$-algebras.
Moreover, let $\tau^1$ be an $\mathcal{F}^1_t-$stopping time and
$\tau^2$ an $\mathcal{F}^2_t-$stopping time.
Now, let us fix a path $(X_t(\omega))_{t\geq 0}$, then conditioned on this path the two stopping times $\tau^1(\omega)\in [0,T]$ and $\tau^2(\omega)\in [0,T]$ can be compared as real-numbers.

Additionally, still conditioned on $(X_t(\omega))_{t\geq 0}$, we may compare the stopped values of a stochastic process $Y_t$, i.e.,
$Y_{\tau^1(\omega)}$ and $Y_{\tau^2(\omega)}$.

\paragraph*{Deep Decision Classifier.}
Given a stopping time $\tau$ achieving \eqref{eq:optStopping} we define the continuation and exercise region at time $t$ by
\begin{align*}
    \mathcal{E}_t^\tau \coloneqq \left\{ X_t(\omega) : \tau(\omega)=t \right\}, \quad
    \mathcal{C}_t^\tau \coloneqq \left\{ X_t(\omega) : \tau(\omega)> t \right\},
\end{align*}
These two sets are naturally labeled sets allowing for supervised learning.

Now, for each point $t$, we would like to find a function $f_t^\tau:\R^d \rightarrow [0,1]$, associated to the probability of value $x\in \R^d$ being in $\mathcal{E}_t^\tau$ and thus with probability $1-f_t(x)$ in $\mathcal{C}_t^\tau$. 
For this classification problem we use at each point in time a Deep-Neural-Network (DNN) for the function $f_t$, whose last activation function is a $\softmax$, using the categorical crossentropy as a loss function.

For the training of the networks, we sample one million paths of the commodities, determine the optimal stopping time by e.g. LSMC, and compute the training sets $\mathcal{E}^\tau_t$ and 
$\mathcal{C}^\tau_t$ at each point in time $t$. Usually, the number of elements in $\mathcal{C}_t^\tau$ is much higher than in $\mathcal{E}^\tau_t$. Hence, we reuse the elements in $\mathcal{E}^\tau_t$ for the training of the network, such that for each batch in the training set we have equal amount of continuations and exercises. This was crucial for the training of the neural networks. The specific hyperparameters and model architecture used in this paper are given in \Cref{tab:dnnParameters} and we leave an optimization of these parameters to the reader. The training for each network takes about $2$ seconds on a CPU in the case $d=2$ and $4$ seconds in the case $d=4$. A GPU did not speed up the training.

\begin{table}[]
    \centering
    \caption{Hyperparameters and architecture of $f_t^\tau$}
    \begin{tabular}{lc}
        Layer/Activation 1 & Batch normalization/ - \\
        Layer/Activation 2 & Dense with $d*16$ neurons/ \relu\\
        Layer/Activation 3 & Dense with $d*32$ neurons/ \relu\\
        Layer/Activation 4 & Dense with $d*16$ neurons/ \relu\\
        Layer/Activation 5 & Batch normalization/ - \\
        Layer/Activation 6 & Dense with $2$ neurons/ \softmax\\
        Batch size &  128 elements of each $\mathcal{E}_t^\tau$ and $\mathcal{C}_t^\tau$\\
        Epochs & Chosen such that the entire training data is used\\
        Optimizer  & Adam with constant learning rate $0.001$ \\
    \end{tabular}
    \label{tab:dnnParameters}
\end{table}

Having the trained neural networks for each point $t$, we can solve \eqref{eq:optStopping} by using these stopping rules, i.e. if 
$f_t^\tau(x)\geq 0.5$ exercise, otherwise continue. We will denote the value of the objective using this rule by $V_0(f^\tau)$. 
Now, having two different stopping times $\tau^1$ and $\tau^2$ we can compare the values $V_0(f^{\tau^1})$ and $V_0(f^{\tau^2})$
with each other. The neural networks have a practical application, since after training one can ask the neural network in the real world scenario whether one should exercise or not at each specific point in time. This is not possible with just using the LSMC. 

\begin{remark}\label{rem:SVM}
    In \citet{Ewald2017}, a simple curve fitting for finding the exercise boundary for the case of constant feeding costs was employed. However, the choice of the regression function and the fit led to a lot of uncertainty and large confidence intervals surrounding the exercise boundary. This is not the case for our DNN. In higher dimensions, for example $d=4$, as in the case of stochastic feeding costs, a curve fitting process is very difficult to perform, if not impossible, since a good intuition for choosing a regression function is hard to obtain.

    As an alternative, one could also use Support-Vector-Machines (SVM) for this classification problem. In fact, we did this as well and observed in the case of deterministic feeding costs reasonably good results, i.e., the inferred stopping rules $f_t$ produced values close to the LSMC value of the objective. 
    However, in the case of stochastic feeding costs the SVMs were not able to classify the exercise values properly. Therefore, we resorted to DNNs, which worked very well.

    Our DNNs have to be understood as a means to compare different stopping rules with each other, however we did not optimise these DNNs for performance. Also they depend on another numerical algorithm to produce the exercise and continuation sets and therefore cannot be used as an alternative to infer the stopping region directly. 

    The reader is referred to \citet{Soner2023} for an approach which is using a DNN for finding the (fuzzy-) stopping region directly.
\end{remark}

\subsection{Results}\label{sec:results}
In this section, we will assume stochastic feeding costs and demonstrate the differences and improvements when the stopping rules obtained from accounting for stochastic feeding costs are employed as compared to the case where deterministic feeding costs are assumed.
We compare all the different scenarios using all the combinations of the parameters given in \Cref{tab:paramSalmon}--\ref{tab:paramSoy}. For the validation set and the pathwise comparison of the stopping times, we used $M=100000$ simulations in the LSMC, which takes roughly $5$ seconds on the CPU in the case of stochastic feeding costs. We used polynomials up to degree two as a regression basis. 

In \Cref{tab:relImprovements}, we show two metrics for the relative improvement of using stochastic feeding costs compared to deterministic feeding costs. For this we use the pathwise comparison and the one using Deep Classifiers, denoted by $\relImprovement^\tau\coloneqq \frac{V_0^{\tau,\text{stoch}}}{V_0^{\tau,\text{determ}}}$ and $\relImprovement^f\coloneqq \frac{V_0^{f,\text{stoch}}}{V_0^{f,\text{determ}}}$, respectively. 

As expected, we notice in the case of \emph{low vol} (first row of \Cref{tab:relImprovements}) that the benefit of using a stopping rule accounting for the stochastic nature of the feed costs  is only slightly better than its deterministic counterpart. In this case deterministic feeding costs seem to be a good approximation.
The benefit of using stochastic feeding costs however increases up to $11.6\,\%$ the higher the volatility, for the scenario of declining salmon prices. 

The DNNs $f_t$ produce close values compared to the pathwise comparison in both \Cref{tab:relImprovements} and \Cref{tab:farmValue}, showing that the neural networks are able to classify the stopping region very well.

In \Cref{tab:farmValue}, we present the corresponding values of the fish farm \eqref{eq:objective} using the different stopping rules for all of the scenario combinations. Further, we stress how much these values change in relation to changes in the risk-premium $\lambda^1$, showing that the decision-making in the real option is very sensitive to the model parameters.

\begin{table}[]
    \caption{Relative improvements using stochastic feeding costs.}
    \centering
    \renewcommand{\arraystretch}{1.5}
    \begin{tabularx}{\linewidth}{c|*{3}{X}}
        \diagbox{Soy}{Salmon}   & down, down & down, up & up, up\\
        \hline
        low vol                 & $\relImprovement^\tau=1.005$\newline
                                  $\relImprovement^f=1.004$
                                & $\relImprovement^\tau=1.002$\newline
                                  $\relImprovement^f=1.002$
                                & $\relImprovement^\tau=1.00019$\newline
                                  $\relImprovement^f=1.00007$
                                \\
        medium vol              & $\relImprovement^\tau=1.029$\newline
                                  $\relImprovement^f=1.029$
                                & $\relImprovement^\tau=1.019$\newline
                                  $\relImprovement^f=1.018$
                                & $\relImprovement^\tau=1.005$\newline
                                  $\relImprovement^f=1.004$
                                \\
        high vol                & $\relImprovement^\tau=1.116$\newline
                                  $\relImprovement^f=1.116$
                                & $\relImprovement^\tau=1.088$\newline
                                  $\relImprovement^f=1.089$
                                & $\relImprovement^\tau=1.045$\newline
                                  $\relImprovement^f=1.045$
    \end{tabularx}
    \label{tab:relImprovements}
\end{table}

\begin{table}[]
    \caption{Value of \eqref{eq:objective} using stochastic feeding costs with different stopping rules.}
    \centering
    \renewcommand{\arraystretch}{1.5}
    \begin{tabularx}{\linewidth}{c|*{3}{X}}
        \diagbox{Soy}{Salmon}   & down, down & down, up & up, up\\
        \hline
        low vol                 & $V_0^{\tau,\text{determ}}=1920575$\newline
                                  $V_0^{\tau,\text{stoch}}=1929700$\newline
                                  $V_0^{f,\text{determ}}=1920498$\newline
                                  $V_0^{f,\text{stoch}}=1928452$
                                & $V_0^{\tau,\text{determ}}=2395735$\newline
                                  $V_0^{\tau,\text{stoch}}=2401140$\newline
                                  $V_0^{f,\text{determ}}=2395467$\newline
                                  $V_0^{f,\text{stoch}}=2399075$
                                & $V_0^{\tau,\text{determ}}=3897624$\newline
                                  $V_0^{\tau,\text{stoch}}=3898397$\newline
                                  $V_0^{f,\text{determ}}=3896446$\newline
                                  $V_0^{f,\text{stoch}}=3896728$
                                \\
        medium vol              & $V_0^{\tau,\text{determ}}=1947469$\newline
                                  $V_0^{\tau,\text{stoch}}=2005608$\newline
                                  $V_0^{f,\text{determ}}=1947200$\newline
                                  $V_0^{f,\text{stoch}}=2003756$
                                & $V_0^{\tau,\text{determ}}=2426319$\newline
                                  $V_0^{\tau,\text{stoch}}=2471621$\newline
                                  $V_0^{f,\text{determ}}=2425987$\newline
                                  $V_0^{f,\text{stoch}}=2469014$
                                & $V_0^{\tau,\text{determ}}=3931880$\newline
                                  $V_0^{\tau,\text{stoch}}=3950537$\newline
                                  $V_0^{f,\text{determ}}=3930926$\newline
                                  $V_0^{f,\text{stoch}}=3946699$
                                \\
        high vol                & $V_0^{\tau,\text{determ}}=2001673$\newline
                                  $V_0^{\tau,\text{stoch}}=2232893$\newline
                                  $V_0^{f,\text{determ}}=2001876$\newline
                                  $V_0^{f,\text{stoch}}=2234064$
                                & $V_0^{\tau,\text{determ}}=2488125$\newline
                                  $V_0^{\tau,\text{stoch}}=2706428$\newline
                                  $V_0^{f,\text{determ}}=2482972$\newline
                                  $V_0^{f,\text{stoch}}=2704694$
                                & $V_0^{\tau,\text{determ}}=3997579$\newline
                                  $V_0^{\tau,\text{stoch}}=4179171$\newline
                                  $V_0^{f,\text{determ}}=3992845$\newline
                                  $V_0^{f,\text{stoch}}=4173701$
    \end{tabularx}
    \label{tab:farmValue}
\end{table} 
Let us focus on the first column in \Cref{tab:farmValue}, the \emph{down, down} scenario. The values $V_0^{\tau,\text{determ}}$ are close too each other regardless of the volatility scenario for the feeding costs. On the other hand, the values $V_0^{\tau,\text{stoch}}$ differ significantly and increase with higher volatility.
This means if the model is misspecified with e.g. too low volatility, then the deterministic stopping rules will produce less revenue, than the stochastic stopping rules, which can benefit from the higher volatility. To mitigate the model misspecification, the neural networks $f_t$ should be trained for different volatilities, such that a unified decision rule can be applied for all cases. This improvement as well as using recurrent neural networks for $f_t$ instead of different neural networks for each $t$ will be left for future research.

If the initial feeding costs would have been $50\,\%$ of the production costs, then e.g. in the case \emph{down, up} with \emph{medium vol} the benefit of using stochastic feeding costs would increase to roughly $8\,\%$. This is similar for all the other cases as well and highlights the importance of determining the correct relationship of salmon revenue and feeding costs. 
For feeding costs between $25\,\%$ and $50\,\%$ of the production costs the benefit will lie in between $1.5\,\%$ and $8\,\%$. Thus, the benefits presented in \Cref{tab:relImprovements} can be understood as a lower bound.
\section{Conclusion}\label{sec:conclusion}
In this article, we focused on the question of whether accounting for stochastic feeding costs can 
significantly improve aquaculture decision making and increase the financial value of a fish farm 
as compared to the case when deterministic feeding costs are assumed. We found a partial answer: If the volatility of the feed price is high enough, then it is beneficial to use stochastic feeding costs, otherwise deterministic feeding costs can lead to an almost equal performance. In none of the cases, however, does the inclusion of stochastic feeding costs lead to impaired performance, and as the computational effort only marginally increases when using our proposed methodology, we highly recommend to adopt relevant models to account for feeding cost risk. In addition to that, as we demonstrated, a further advantage is that the inclusion of stochastic feeding costs can make the optimal stopping rule more robust with respect to model misspecification.

We found that there is a lot of model uncertainty, highlighted by the significantly different parameters obtained from Kalman filtering and the technique by \citet{Cortazar2003}. Since this has a major impact on the real option, both value and exercise, it needs to be addressed in future work. To this end we would like to add a control to the objective \eqref{eq:optStopping} in terms of a model-free decision rule, based on so-called myopic look ahead and $k$-step ahead rules (cf. \citet{Ross1971}). It might be possible to add such a control also to the Kalman filter allowing for more robust parameter estimations.

In the line of feeding costs, we have seen that the relative difference of the salmon prices to the feeding costs is important. If salmon prices are much higher/ lower than feeding costs and tend to increase/ decrease a lot, then the stochastic nature of the feeding costs may be negligible/ dominating.
To investigate this further, we could also study a more realistic case, in which storage for the feed is available, as well as optimal pairs of buying time and buying quantity combined with an adjustable feeding rate per day. Another opportunity for future research is to further investigate the correlation between salmon price and salmon feed and how this should be implemented and estimated in classical multi-factor commodity models. An application of reinforcement learning comes to mind to tackle this problem.

Moreover, we are interested in a case where the inter-correlations of the commodities may be stochastic as well. As this leads to a full correlation matrix, it can be modeled with a stochastic process taking values in an appropriate Lie-group like in \citet{Muniz2021}. For this the techniques developed in \citet{Muniz2022} and \citet{KPP2021} may be very helpful.

In \citet[pp.~40--45]{Misund2022} it is argued that biological costs due to e.g. salmon lice treatments are of equal size to the feeding costs. Thus, in future studies, one should address a similar question for stochastic mortality or at an even deeper level of detail, look at how host-parasite models and costs for salmon lice treatments can be implemented with the classical salmon farm model. 
\appendix

\section*{Declarations}
The authors have no relevant financial or non-financial interests to disclose.
\vspace{-5mm}
\printbibliography

@article{Schwartz1997,
author = {Schwartz, Eduardo S.},
title = {The Stochastic Behavior of Commodity Prices: Implications for Valuation and Hedging},
journal = {The Journal of Finance},
volume = {52},
number = {3},
pages = {923-973},
doi = {10.1111/j.1540-6261.1997.tb02721.x},
year = {1997}
}

@article{Ewald2017,
author = {Ewald, Christian-Oliver and Ouyang, Ruolan and Siu, Tak Kuen},
title = {On the Market-consistent Valuation of Fish Farms: Using the Real Option Approach and Salmon Futures},
journal = {American Journal of Agricultural Economics},
volume = {99},
number = {1},
pages = {207-224},
doi = {10.1093/ajae/aaw052},
year = {2017}
}

@article{Cortazar2003,
title = {Implementing a stochastic model for oil futures prices},
journal = {Energy Economics},
volume = {25},
number = {3},
pages = {215-238},
year = {2003},
issn = {0140-9883},
doi = {https://doi.org/10.1016/S0140-9883(02)00096-8},
author = {Gonzalo Cortazar and Eduardo S. Schwartz}
}

@article{Becker2021,
doi = {10.1017/s0956792521000073},
year = 2021,
month = {apr},
publisher = {Cambridge University Press ({CUP})},
volume = {32},
number = {3},
pages = {470--514},
author = {Sebastian Becker and Patrick Cheridito and Arnulf Jentzen and Timo Welti},
title = {Solving high-dimensional optimal stopping problems using deep learning},
journal = {European Journal of Applied Mathematics}
}

@article{Longstaff2001,
title = {Valuing American Options by Simulation: A Simple Least-Squares Approach},
author = {Longstaff, Francis and Schwartz, Eduardo S},
year = {2001},
journal = {Review of Financial Studies},
volume = {14},
number = {1},
pages = {113--147},
doi = {10.1093/rfs/14.1.113},
}

@misc{Soner2023,
title={Neural Optimal Stopping Boundary}, 
author={A. Max Reppen and H. Mete Soner and Valentin Tissot-Daguette},
year={2023},
eprint={2205.04595},
archivePrefix={arXiv},
primaryClass={q-fin.PR}
}

@Article{Muniz2021,
AUTHOR = {Muniz, Michelle and Ehrhardt, Matthias and Günther, Michael},
TITLE = {Approximating Correlation Matrices Using Stochastic Lie Group Methods},
JOURNAL = {Mathematics},
VOLUME = {9},
YEAR = {2021},
NUMBER = {1},
ARTICLE-NUMBER = {94},
ISSN = {2227-7390},
DOI = {10.3390/math9010094}
}

@Article{Muniz2022,
author={Muniz, Michelle
and Ehrhardt, Matthias
and G{\"u}nther, Michael
and Winkler, Renate},
title={Correction to: Higher strong order methods for linear It{\^o} SDEs on matrix Lie groups},
journal={BIT Numerical Mathematics},
year={2022},
month={Sep},
day={01},
volume={62},
number={3},
pages={1093-1093},
issn={1572-9125},
doi={10.1007/s10543-022-00911-5},
}

@Article{KPP2021,
author={Kamm, Kevin
and Pagliarani, Stefano
and Pascucci, Andrea},
title={On the Stochastic Magnus Expansion and Its Application to SPDEs},
journal={Journal of Scientific Computing},
year={2021},
month={Oct},
day={18},
volume={89},
number={3},
pages={56},
issn={1573-7691},
doi={10.1007/s10915-021-01633-6},
}

@Article{Misund2022,
author={Misund, Bard},
title={Cost Development In Atlantic Salmon and Rainbow Trout Farming: What Is The Cost of Biological Risk? },
journal={SSRN},
year={2022},
month={Dec},
day={19},
doi={10.2139/ssrn.4307278},
}

@Book{Bjork2009,
author={Bj{\"o}rk, Tomas},
title={Arbitrage theory in continuous time},
series={Oxford finance},
year={2009},
edition={3. ed.},
publisher={Oxford University Press},
address={Oxford ;},
note={Includes bibliographical references and index},
isbn={9780199574742}
}

@Book{Harvey1989,
author={Harvey, Andrew C.},
title={Forecasting, structural time series models and the Kalman filter},
year={1989},
publisher={Cambridge University Press},
address={Cambridge},
isbn={0521321964}
}

@book{Bishop2006,
author = {Bishop, Christopher M.},
title = {Pattern Recognition and Machine Learning (Information Science and Statistics)},
year = {2006},
isbn = {0387310738},
publisher = {Springer-Verlag},
address = {Berlin, Heidelberg}
}

@article{Shepherd2017,
title = {Future availability of raw materials for salmon feeds and supply chain implications: The case of Scottish farmed salmon},
journal = {Aquaculture},
volume = {467},
pages = {49-62},
year = {2017},
note = {Cutting Edge Science in Aquaculture 2015},
issn = {0044-8486},
doi = {https://doi.org/10.1016/j.aquaculture.2016.08.021},
author = {C. Jonathan Shepherd and Oscar Monroig and Douglas R. Tocher}
}

@article{Gomes2023,
title = {Statistical modelling of voluntary feed intake in individual Atlantic salmon (Salmo salar L.)},
year = {2023},
journal={Frontiers in Marine Science},
doi = {10.3389/fmars.2023.1127519},
author = {Gomes, AS and Zimmermann, F and Hevrøy, EM and Søyland, MAL and Hansen, TJ and Nilsen, TO and Rønnestad, I}
}

@Article{Osmundsen2021,
AUTHOR = {Osmundsen, Kjartan Kloster and Kleppe, Tore Selland and Liesenfeld, Roman and Oglend, Atle},
TITLE = {Estimating the Competitive Storage Model with Stochastic Trends in Commodity Prices},
JOURNAL = {Econometrics},
VOLUME = {9},
YEAR = {2021},
NUMBER = {4},
ARTICLE-NUMBER = {40},
ISSN = {2225-1146},
DOI = {10.3390/econometrics9040040}
}

@article{Luna2023,
title = {A conceptual framework for risk management in aquaculture},
journal = {Marine Policy},
volume = {147},
pages = {105377},
year = {2023},
issn = {0308-597X},
doi = {10.1016/j.marpol.2022.105377},
author = {Manuel Luna and Ignacio Llorente and Ladislao Luna}
}

@article{Ross1971,
 ISSN = {00034851},
 URL = {http://www.jstor.org/stable/2958480},
 author = {Sheldon M. Ross},
 journal = {The Annals of Mathematical Statistics},
 number = {1},
 pages = {297--303},
 publisher = {Institute of Mathematical Statistics},
 title = {Infinitesimal Look-Ahead Stopping Rules},
 volume = {42},
 year = {1971}
}

\end{document}